\documentclass[osajnl,twocolumn,showpacs,superscriptaddress,10pt]{revtex4-1} %% use 11pt for Applied Optics
\usepackage{amsmath,amssymb,graphicx}

\usepackage{mathtools}
\usepackage{xfrac}
\begin{document}

\title{Modelocked mid-infrared frequency combs in a silicon microresonator}

\author{Mengjie Yu}\email{Corresponding author: my2473@columbia.edu}
\affiliation{Department of Applied Physics and Applied Mathematics, Columbia University, New York, NY 10027}
\affiliation{School of Electrical and Computer Engineering, Cornell University, Ithaca, NY 14853}

\author{Yoshitomo Okawachi}
\affiliation{Department of Applied Physics and Applied Mathematics, Columbia University, New York, NY 10027}

\author{Austin G. Griffith}
\affiliation{School of Applied and Engineering Physics, Cornell University, Ithaca, NY 14853}

\author{Michal Lipson}
\affiliation{Department of Electrical Engineering, Columbia University, New York, NY 10027}

\author{Alexander L. Gaeta}
\affiliation{Department of Applied Physics and Applied Mathematics, Columbia University, New York, NY 10027}

\begin{abstract}Mid-infrared (mid-IR) frequency combs have broad applications in molecular spectroscopy and chemical/biological sensing. Recently developed microresonator-based combs in this wavelength regime could enable portable and robust devices using a single-frequency pump field. Here, we report the first demonstration of a modelocked microresonator-based frequency comb in the mid-IR spanning 2.4 \textmu m to 4.3 \textmu m.  We observe high pump-to-comb conversion efficiency, in which 40$\%$ of the pump power is converted to the output comb power. Utilizing an integrated PIN structure allows for tuning the silicon microresonator and controling modelocking and cavity soliton formation, simplifying the generation, monitoring and stabilization of mid-IR frequency combs via free-carrier detection and control. Our results significantly advance microresonator-based  comb technology towards a portable and robust mid-IR spectroscopic device that operates at low pump powers.
\end{abstract}

\ocis{(140.3948) Microcavity devices; (190.4975) Parametric processes; (190.4390) Integrated optics}% REPLACE WITH CORRECT OCIS CODES FOR YOUR ARTICLE
                          % NOTE: \ocis{} IS ALIASED TO \pacs{} BUT MUST
                          % FORMAT THE TERMS CORRECTLY FOR EACH JOURNAL

\maketitle %% required

\section{Introduction}

\begin{figure*}[t!]
\centering{\includegraphics[width=\linewidth]{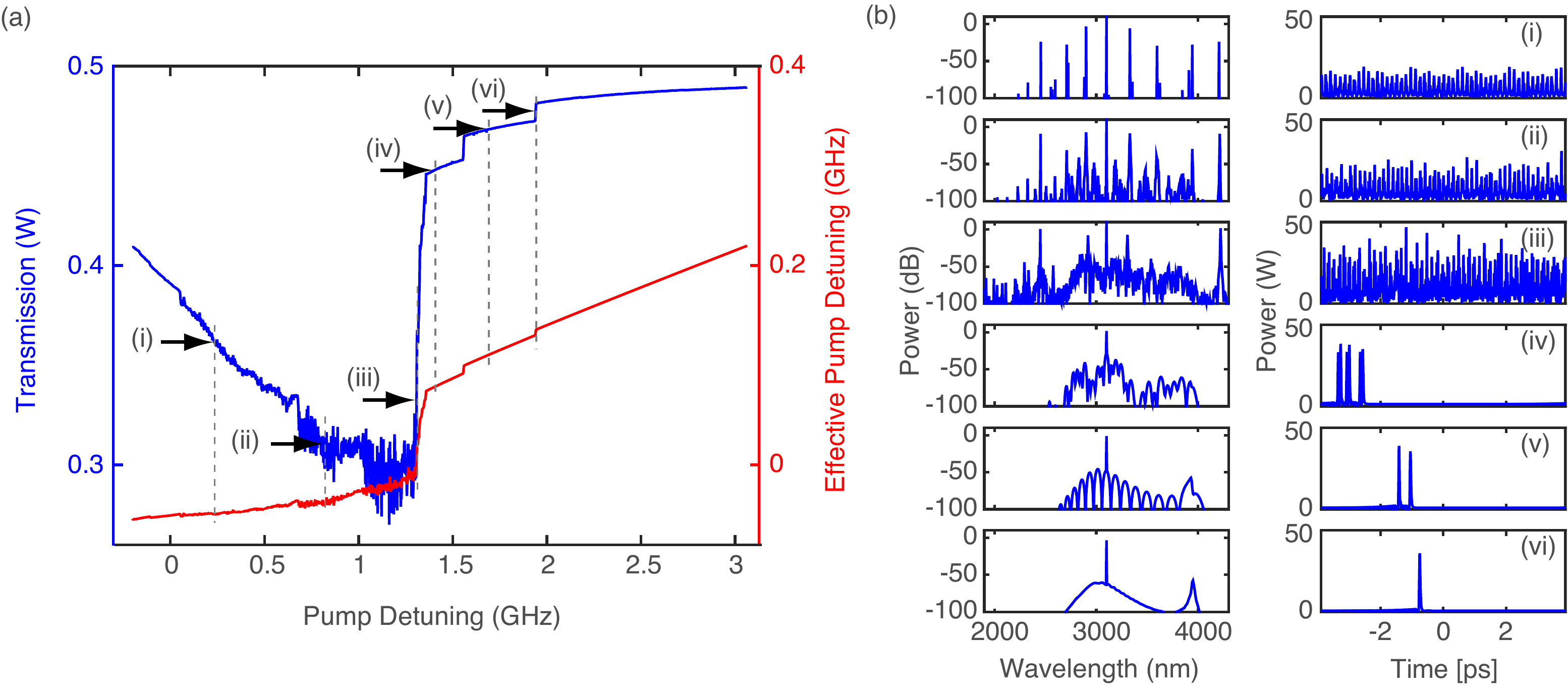}}
%\fbox{\includegraphics[width=\linewidth]{Fig1}}
\caption{Numerical simulation of soliton formation via pump laser detuning in a silicon microresonator. a) Transmission and effective pump-cavity detuning when scanning a pump laser over a cavity resonance. Effective pump-cavity detuning subtracts the resonance shift due to the Kerr effect and free-carrier dispersion from pump detuning. A series of abrupt transmission steps occur in the effective red-detuning regime. b) Optical spectra and intracavity temporal behavior at different positions (i-vi) in the scan. The plots associated with (i-iii) are in the effective blue-detuning regime while plots (iv-vi) are in the effective red-detuning regime. The transition to modelocking and soliton formation occurs between (iii) and (iv), as the pump frequency is tuned across the zero effective pump detuning point. As the laser is tuned, (i) primary sidebands are generated through modulation instability and cascaded FWM, (ii) mini-combs are generated around each primary sideband, (iii) with further intracavity power building up, different sets of mini-combs overlap producing high-noise temporal behavior, (iv) a multiple (6) soliton state occurs with a complex, structured optical spectrum, (v) a 2-soliton state is produced, and (vi) the single soliton state with a smooth optical spectrum is reached. States with a different number of solitons correspond to the transmission steps shown in (a).}
\label{Fig1}
\end{figure*}

Optical frequency comb generation using four-wave mixing (FWM) parametric oscillation in microresonators has attracted significant interest \cite{Kippenberg, Levy, Okawachi, Herr12, Jung, Savchenkov, Papp, Liu, Herr, Hausmann, Huang,Brasch}. Progress in fabrication techniques for low-loss nonlinear devices has lead to significant development of comb generation in the mid-infrared (mid-IR) regime \cite{Luke,Griffith,GriffithRaman,Shankar,CWang,Savchenkov15, Lecaplain}, which is a highly attractive spectral region for applications in molecular spectroscopy and chemical and biological sensing \cite{Schliesser}. For example, broadband mid-IR sources, especially from 3 to 4 \textmu m, are useful for breath analysis, since many different species are exhaled, such as carbonyl sulfide, ethane, ethylene and formaldehyde, which are important biomarkers for various diseases. Mid-IR parametric oscillation and frequency comb generation have been demonstrated in silicon nitride (Si$_3$N$_4$) \cite{Luke}, silicon \cite{Griffith, GriffithRaman} microresonators, and crystalline calcium fluoride and magnesium fluoride resonators \cite{CWang,Savchenkov15, Lecaplain}. Various spectroscopic techniques could be enabled by broadband coherent comb sources, including multi-heterodyne and dual-comb spectroscopy that yield high sensitivity and fast acquisition speeds \cite{Coddington, Bernhardt, Adler}. 

A critical feature of microresonator-based combs is that they can be modelocked, which results in the comb spacing to be highly uniform across the entire comb. Recently, there have been several demonstrations of such modelocking in the near-infrared regime \cite{Herr, FosterArXiv, Saha, HerrPRL, Yi, Brasch, Karpov}. To achieve modelocking, the microresonators are pumped in the anomalous group-velocity dispersion (GVD) regime, and soliton formation is controlled via pump frequency detuning with respect to the cavity resonance. Several additional tuning methods have been demonstrated, including pump power modulation \cite{Brasch}, electro-optic tuning based on the Pockels effect \cite{Jung14}, and thermal tuning of the cavity resonance \cite{Joshi}, which allow for stable, systematic generation of single-soliton states. Soliton formation techniques not involving pump frequency tuning are particularly attractive, since it would allow for the use of ultra-narrow linewidth frequency-stabilized single-frequency pump sources for comb generation and enable simultaneous generation of frequency combs in multiple rings using a single pump source, which would faciliate applications such as chip-scale dual-comb spectroscopy \cite{Zhu}.

In this paper, we present the first demonstration of coherent mid-IR frequency comb generation via soliton modelocking in silicon microresonators. The evolution towards temporal cavity soliton formation and modelocking is numerically simulated including effects of multiphoton absorption (MPA), free-carrier dispersion (FCD), and free-carrier absorption (FCA). Experimentally, the dynamics are carefully characterized by comb transmission and three-photon absorption (3PA)-induced photocurrent measurements. Modelocking is found to be reproducible at different pump wavelengths and in multiple silicon devices and is repeatedly achievable by appropriate tuning of the pump laser frequency. A 40 $\%$ power conversion from the pump to the modelocked comb is achieved. Furthermore, we demonstrate full control of the modelocking dynamics through electrical tuning of the free-carrier (FC) lifetime, allowing for fixed-pump-frequency operation. Our results show that the distinct features of silicon, which are high nonlinearity, high thermal conductivity, MPA and the FC-induced effect, can be suitably exploited for modelocked broadband mid-IR frequency comb generation.

\begin{figure*}[t!]
\centering{\includegraphics[width=\linewidth]{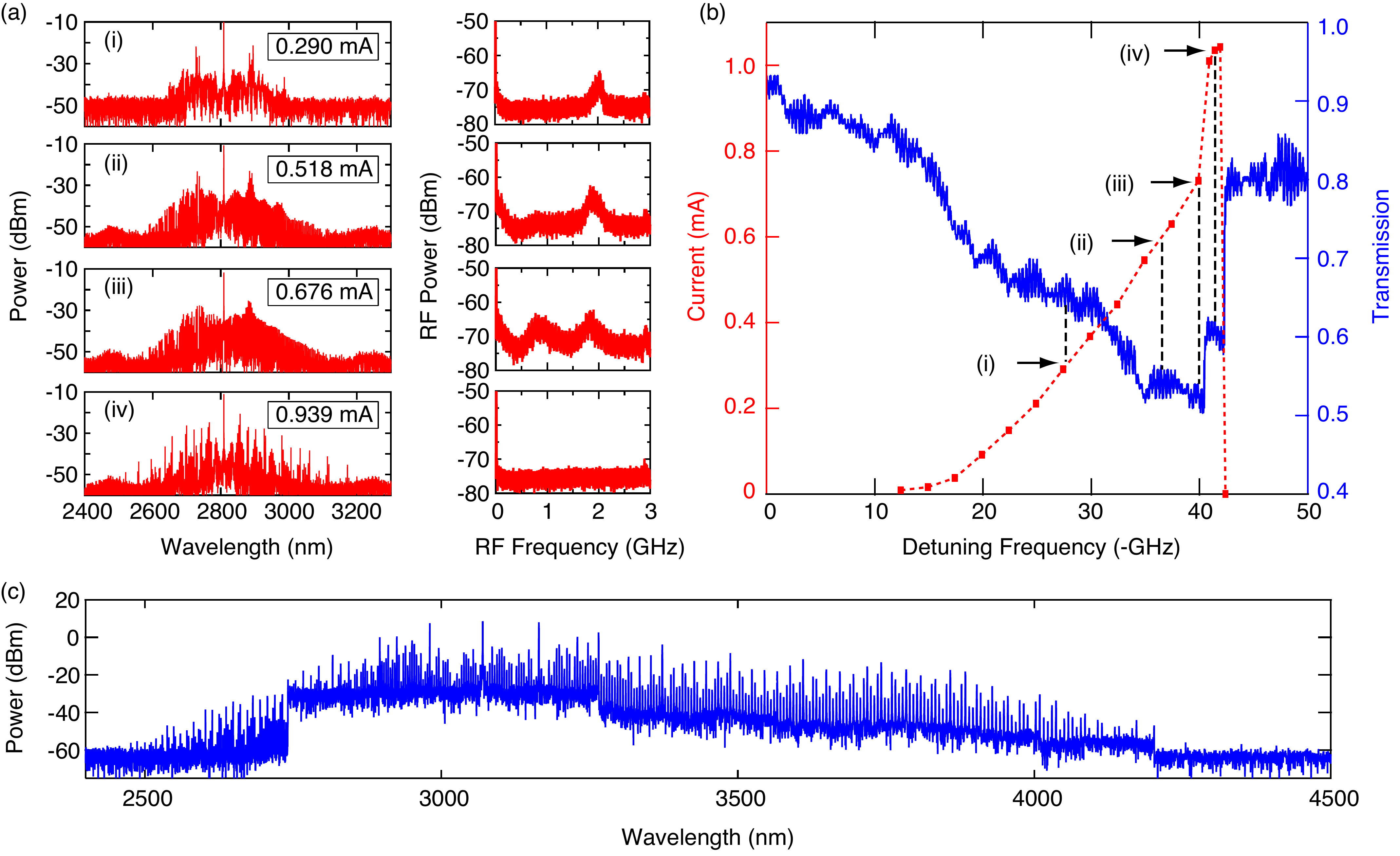}}
\caption{The evolution towards modelocking via pump frequency detuning from the cavity resonance. (a) Left: Optical spectrum. Right: RF spectrum of the 3PA-induced current. (b) Transmission (blue solid curve) and 3PA-induced current (red dashed curve) as the pump frequency is scanned. Arrows correspond to four states (i-iv) in (a). (c) Optical spectrum of soliton mid-IR frequency comb pumped at 3.07 \textmu m.}
\label{Fig2}
\end{figure*}

\section{Modelocking via Pump Detuning}

\subsection{Numerical Simulations}
We carry out numerical simulations on the nonlinear dynamics of comb generation in silicon microresonators using a modified Lugiato-Lefever model \cite{Lugiato,Haelterman,Matsko,Chembo,Coen,Lamont,Wabnitz,Lau}. Based on the device structure of the silicon microresonator described in \cite{Griffith}, the simulated GVD is anomalous (-102 ps$^2$/km) for a fundamental TE mode at a pump wavelength of 3.1 \textmu m. Higher-order dispersion, self-steepening, 3PA, and the FC effect (FC lifetime $\tau_{fc}$=100-ps) are included in the model. Figure \ref{Fig1}(a) shows the optical transmission and the effective pump-cavity detuning as functions of pump detuning from the cold cavity resonance. The effective pump-cavity detuning is calculated by subtracting self-phase modulation, cross-phase modulation and the FC-induced phase shift from the pump detuning from a cold-cavity resonance. For our simulations, we operate in the undercoupled regime. The evolution of the optical spectra and temporal behavior [Fig. \ref{Fig1}(b)] indicates that the soliton state occurs after a transition from a high-noise state and results in high peak power pulses at red effective pump detunings, which coincides with the transmission steps and is consistent with previous observations in other platforms \cite{Herr,Yi,Joshi}. 

Our simulations indicate that the soliton states are achieved at high optical transmission, corresponding to values close to the off-resonance transmission. This condition is attributed to high pump power and low pump-to-comb conversion efficiency, which is defined as the soliton power divided by the minimum pump power for soliton formation. In practice, this results in a significant change in the intracavity power when the comb undergoes a transition to the soliton state, resulting in a sudden resonance shift due to thermal effects which is not included in this model. This thermal shift can affect the allowable range of pump detunings for stable soliton operation, especially in silicon due to its high thermal conductivity and short heat dissipation time. This issue can be overcome, for a given \emph{Q} factor and free spectral range (FSR) by operating in the over-coupled regime of the resonator with low pump power, which allows for high pump efficiency and minimal change in the intracavity power when transitioning to the soliton state \cite{Yi}. In addition, operation in the over-coupled regime alleviates the requirements of some complex techniques used in Si$_3$N$_4$, MgF$_2$, and silica microresonators to overcome the transient instability of the states, such as two-step protocol \cite{Brasch} or searching for an ideal tuning speed of the pump laser \cite{Herr}.

\subsection{Experimental Demonstration}
We use a high-\textit{Q} etchless silicon microresonator with a silica cladding and an integrated PIN structure, as described in \cite{Griffith}. The microresonator has a FSR of 127 GHz and is dispersion engineered to have anomalous GVD beyond 3 \textmu m for the fundamental TE mode. The PIN junction is operated at reverse-bias to shorten the effective lifetime of the FC's generated by 3PA and is critical for mid-IR frequency comb generation in silicon \cite{Griffith, Lau}. The experimental setup is similar to that shown in \cite{GriffithRaman} in which we pump the silicon microresonator with a continuous wave (cw) optical parametric oscillator (Argos Model 2400). The optical spectrum and both the DC and RF components of the extracted 3PA-induced current are monitored simultaneously using an FTIR, a Keithley Sourcemeter, and an RF spectrum analyzer, respectively. 

A unique property of silicon microresonator frequency combs is that the intracavity dynamics can be characterized by the RF modulation in the PIN junction arising directly from 3PA-induced photocurrent \cite{GriffithRaman}. Any RF modulation of the optical field is imprinted on the FC density and thus on the photocurrent. Therefore, we are able to examine the comb dynamics using the PIN diode, similar to using a conventional photodiode. The design of this PIN-based detector achieves a high bandwidth as compared with conventional photodiodes due to the short FC lifetime. For example, $\tau_{fc}$=12 ps is achievable when at a reverse-bias voltage of -15 V, which yields a bandwidth of 13 GHz for this PIN-based detector. 

The measured loaded \emph{Q} factor of the silicon microresonator is 245,000 at 2.8 \textmu m. The device is overcoupled at 2.8 \textmu m, which we verify by measuring a reduced extinction factor when a reverse-bias voltage is applied on the PIN junction and observing a decreased nonlinear loss due to FCA inside the cavity. We generate frequency combs by tuning the cw pump laser at 2.8 \textmu m into resonance at a reverse-bias voltage of -20 V. The measured threshold power for parametric oscillation is 8 mW, which is consistent with the predicted value based on the \emph{Q} factor \cite{VahalaPth}, and the off-resonance pump power in the bus waveguide for soliton modelocking is 35 mW. The comb generation dynamics as the pump power is increased are shown in Fig. \ref{Fig2}(a). In Fig. \ref{Fig2}(a)(i)-(iii), the pump power builds up inside the microresonator, resulting in an increase in DC current. During this process, primary sidebands are formed through modulation instability and cascaded FWM. With further power buildup, mini-comb formation occurs near each of the primary sidebands and results in a loss of spectral coherence creating multiple broad RF beat notes and high RF amplitude noise [Fig. \ref{Fig2}(a)(i-iii)]. Finally, the frequency comb state abruptly transitions to a low-noise state with a more structured optical spectrum [Fig. \ref{Fig2}(a)(iv)]. More importantly, the transition coincides with an abrupt increase in the DC current from 0.676 mA to 0.939 mA, which is a strong indication of pulse formation inside the cavity, since the 3PA-induced current is proportional to the cube of temporal peak power. In addition, we slowly scan the pump frequency at a 0.1 Hz rate across the resonance while monitoring the transmission using an amplified PbSe detector, as shown in Fig. \ref{Fig2}(b). The transmission shows a clear step, which deviates from the expected triangular resonance shape, due to the large Kerr shift induced by intracavity soliton formation [4] and is well predicted by our simulations (Fig. \ref{Fig1}). Moreover, the position of the transmission step corresponds to the abrupt change in current. The fact that the current increases despite the decrease in intracavity average power, is another strong indication of soliton formation.

As discussed previously for other platforms \cite{Herr,Brasch}, the excitation of stable soliton states is often complicated by the thermal nonlinearity of the resonator. Silicon has a large thermal-optic effect in the mid-IR and a large thermal diffusivity (10 times larger than Si$_3$N$_4$), which makes the thermal instability in the microresonator more detrimental than in other platforms. Despite the thermal effect, the final multi-soliton state can be reached with a slow scanning speed and can be repeatedly achieved by manually tuning the pump laser. This is attributed to operating in the overcoupled regime with a relatively low pump power, where soliton formation can still be achieved due to the large nonlinearity of silicon. Further optimization of the parameters to reduce the transient thermal instability may allow for the observation of multiple soliton steps. In our case, the number of observable steps is limited by the response of the lock-in detection and the scanning speed of the pump laser. 

Next, we obtain a modelocked mid-IR frequency comb with near-octave spanning spectrum spanning 2.4 -- 4.3 \textmu m by pumping at 3.07 \textmu m with 80 mW in the bus waveguide [Fig. \ref{Fig2}(c)]. In the frequency domain, the comb spans from 70 THz to 121 THz, which corresponds to more than 400 modes evenly spaced by 127 GHz. The high wavelength side of the frequency comb is capped at 4.3 \textmu m, which is due to the intrinsic loss in the silica cladding and air absorption (primarily due to CO$_2$). A higher pump power to achieve soliton formation is required, largely due to the slightly lower \emph{Q} factor at 3.07 \textmu m. Unlike previously measured soliton spectra, we observe a significantly depleted pump mode, which we attribute to operating in the highly overcoupled microresonator. The output comb power excluding the pump mode is 32 mW and comprises 80$\%$ of the total output power of 40 mW, which indicates a 40$\%$ pump-to-comb efficiency. To our knowledge, it represents the broadest modelocked frequency comb demonstrated in microresonators.

\begin{figure}[t!]
\centering{\includegraphics[width=\linewidth]{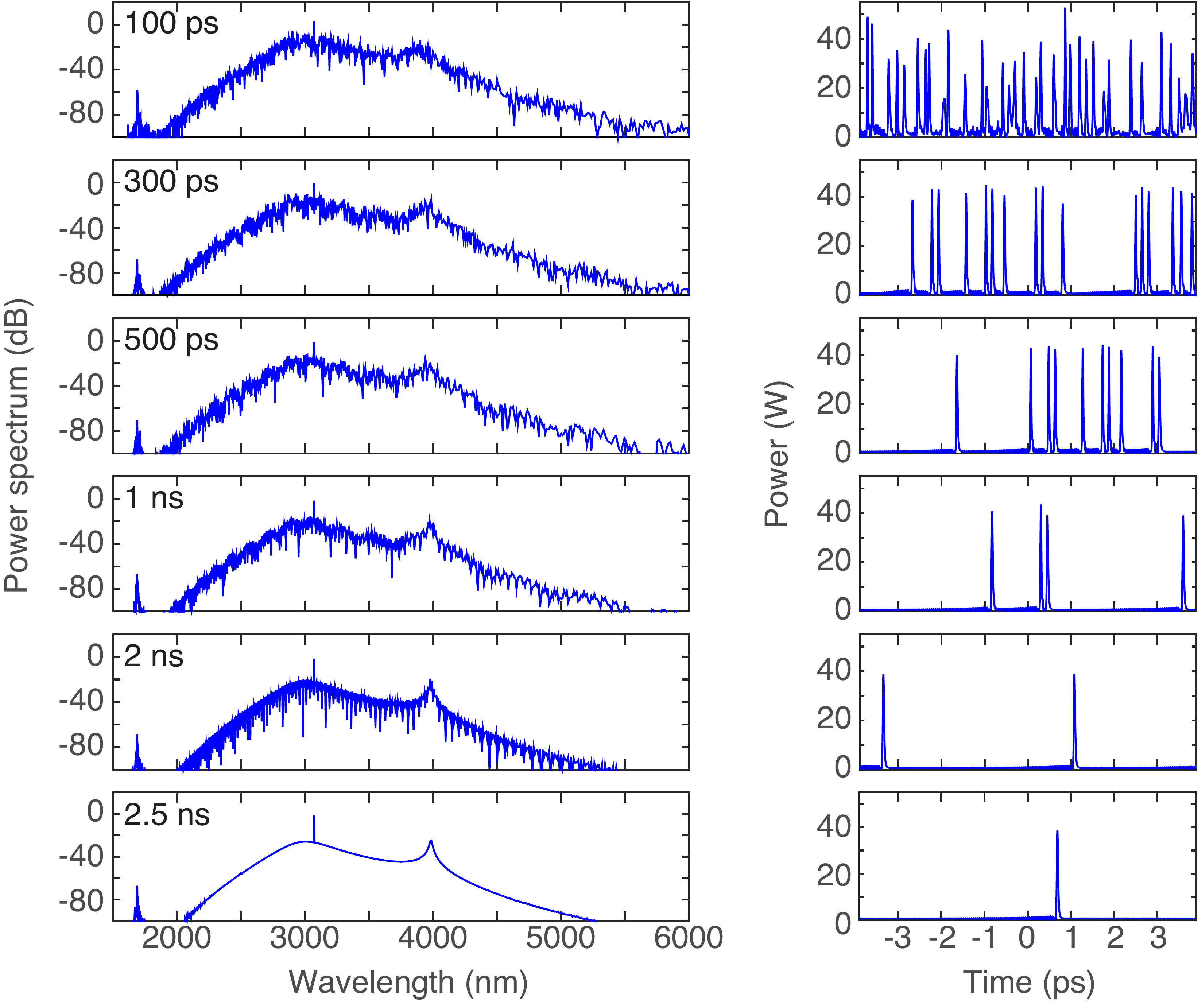}}
\caption{Numerical simulations of soliton formation via free-carrier (FC) lifetime tuning in a silicon microresonator. Starting with a high-noise state frequency comb, the evolution of the optical spectra (left) and temporal behavior (right) is shown by increasing the FC lifetime from 100 ps to 2.5 ns. The full time axis represents one round trip of the microresonator.}
\label{Fig3}
\end{figure}

\begin{figure*}[t!]
\centering{\includegraphics[width=\linewidth]{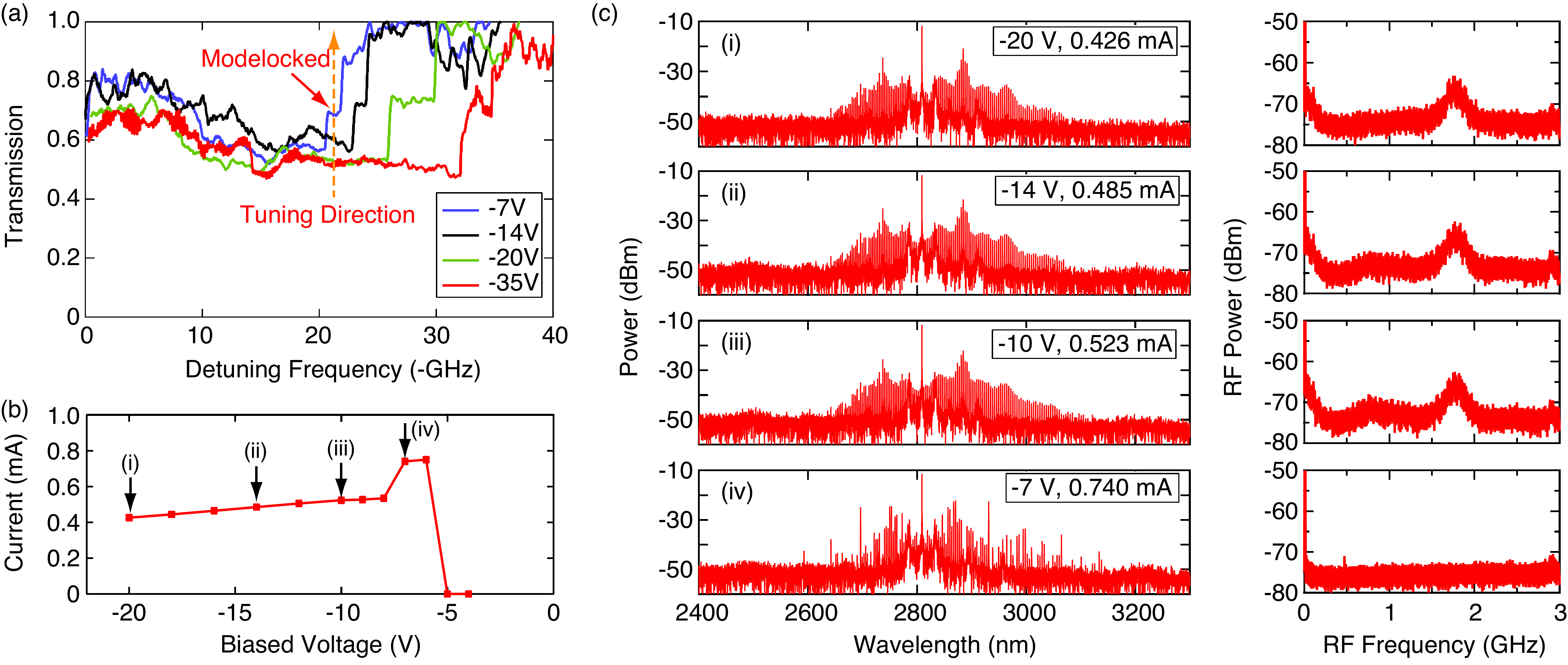}}
\caption{Electrical tuning of FC lifetime towards modelocking. (a) Optical transmission while scanning the pump frequency over one cavity resonance at different reverse-bias voltages. Orange arrow indicates the voltage tuning path to achieve soliton modelocking, corresponding to the generation dynamics in (c). (b) 3PA-induced current following the tuning path in (a). (c) Comb generation dynamics via electrical tuning. Right: optical spectrum; left: RF spectrum. (i-iv) correspond to various reverse-bias values indicated in (b).}
\label{Fig4}
\end{figure*}

\section{Modelocking via Electrical Tuning}
Using pump laser detuning to achieve, control, and stabilize solitons is limiting since tunable lasers typically suffer from frequency jitter, mode hopping, and limited tuning speed. Therefore, in order to enable an ultralow noise frequency comb with long-term stability, an alternative tuning method combined with a fixed frequency pump laser is beneficial. For example, it will allow for interfacing with quantum cascaded lasers (QCL), which would be a significant step towards fully integrated operation and for achieving comb generation in multiple microresonators with a single pump. Thermal tuning of a cavity resonance with an integrated heater has recently been achieved \cite{Weiner}, and Joshi, \emph{et al.} \cite{Joshi} have demonstrated modelocking with this technique. Our approach here is to achieve a modelocked mid-IR frequency comb in silicon by controlling the FC lifetime. This novel approach takes advantage of the negative FCD coefficient and introduces a blue-shifted cavity resonance during comb generation.  In our devices, the FC lifetime can be controlled by changing the applied reverse-bias voltage on the PIN junction, which has been previously reported in \cite{Foster} showing a tuning capability from 3 ns to 12.2 ps for reverse-bias values ranging from 0 to -15 V in a similar PIN junction structure. In addition, since the FC lifetime determines the bandwidth of this tuning method, the tuning speed can be much faster than with thermal tuning. 

\subsection{Numerical Simulation}

We use the modified Lugiato-Lefever model using our silicon device parameters to numerically simulate FC lifetime tuning for modelocking. The 3PA coefficient is 2$\times$10$^{-27}$ m$^3$/W$^2$, and FCA cross section and FCD parameter are 5.88$\times$10$^{-21}$ m$^2$ and 3.75, respectively, as defined in \cite{Lau}. The range of FC lifetimes used in our model is within the experimental capability based on the range of applicable reverse-bias voltages. In our simulations, since FC's due to 3PA plays a role above a certain intracavity pump power, the frequency combs are first generated via regular linear pump frequency detuning at $\tau_{fc}$=100 ps [Fig. \ref{Fig3}(i)], which is followed by an increase in FC lifetime to 300 ps while pump detuning is kept fixed [Fig. \ref{Fig3}(ii)]. This results in a change in pump detuning from effective blue-detuned to red-detuned, along with the evidence of multiple soliton formation. Further increases in FC lifetime from 300 ps to 500 ps, 1 ns and eventually 2 ns reduces the number of solitons in the cavity [Fig. \ref{Fig3}(iii-v)], and the single soliton state with a smooth optical spectrum is achieved at $\tau_{fc}$=2.5 ns [Fig. \ref{Fig3}(vi)]. In addition, our simulations indicate that soliton formation can be achieved at larger pump detuning values for $\tau_{fc}$=100 ps, which is equivalent to the conventional method of pump laser detuning and is due to the fact that such electrical tuning introduces an additional nonlinear cavity detuning via FCD. Also, since FCD relies on the 3PA process, it depends on the dynamical behavior of intracavity power and changes the FCA-induced nonlinear cavity loss at the same time.

\begin{figure*}[t]
\centering{\includegraphics[width=\linewidth]{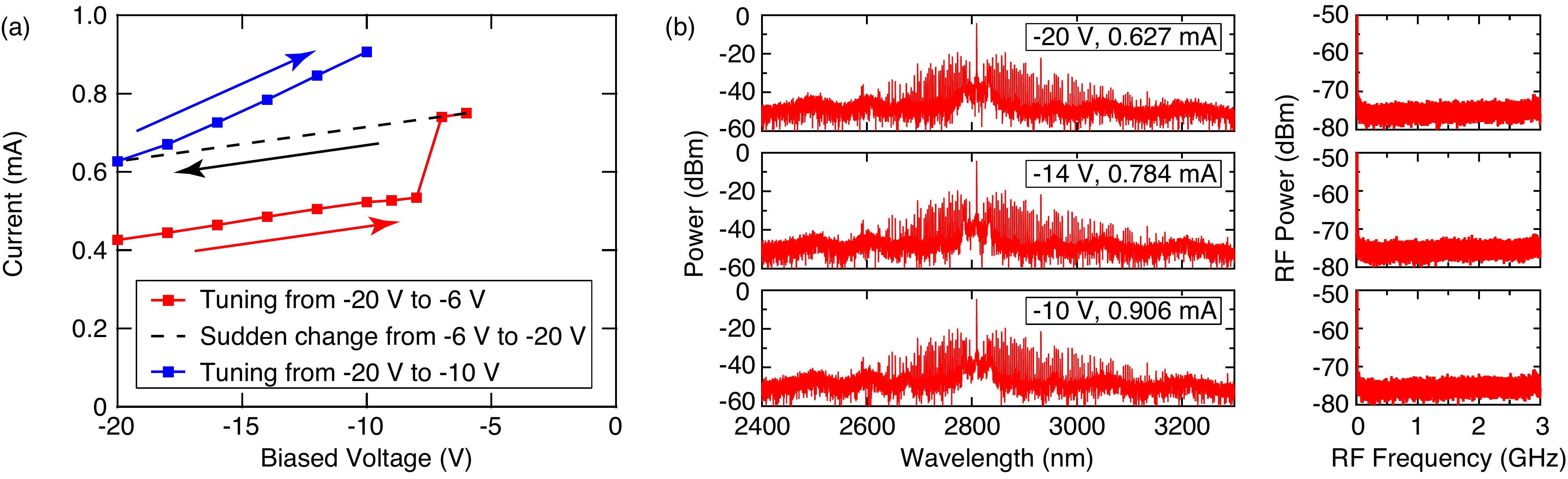}}
\caption{Electrical measurement and control of the soliton state. (a) DC component of 3PA-induced current measured with a specific reverse-bias voltage tuning path. (b) Optical spectra and RF noise spectra of the soliton state at three reverse-bias voltages corresponding to the blue arrow in (a).}
\label{Fig5}
\end{figure*}

\subsection{Experimental Demonstration}
We first investigate the transmission across a cavity resonance by scanning the pump laser frequency for different reverse-bias voltages of -35 V, -20 V, -14 V, and -7 V, as shown in Fig. \ref{Fig4}(a). We see the cavity resonance blue-shifts with decreasing voltage, indicating that we are in the regime where FCD dominates and that, more importantly, tuning the reverse-bias voltage changes the pump detuning from the cavity resonance. Characteristic soliton steps are observed within this voltage range, but not below -6 V due to increased FCA.  To achieve modelocking, we follow the tuning direction, as shown in Fig. \ref{Fig4}(a), from -20 V to -7 V which corresponds to increasing the FC lifetime. Starting from a high-noise state at -20 V, the frequency comb is modelocked at -7 V, corresponding to the position of the soliton step in the transmission. This transition also coincides with an abrupt increase of the 3PA-induced current even at a decreased reverse-bias voltage, which is indicative of intracavity pulse formation [Fig. \ref{Fig4}(b)]. The corresponding evolution with both optical and RF spectra are recorded in Fig. \ref{Fig4}(c). The final multi-soliton state occurs with a modulated optical spectrum and low RF amplitude noise. There exists multiple voltage tuning paths to achieve modelocking. For example, seen from Fig. \ref{Fig4}(a), if the initial reverse-bias voltage is -35 V, the final reverse-bias voltage for modelocking can range from -6 V (the minimum reverse-bias voltage at which modelocking still exists) up to -35 V, depending on the initial pump detuning.

\subsection{Stability of Soliton States}
Once soliton formation is achieved, we explore the robustness of the soliton state by varying the reverse-bias voltage. Once a particular soliton state is reached, the range over which the reverse-bias voltage can be decreased corresponds to the detuning range of the given soliton step, which, in our case, is from -7 V to -6 V. However, soliton states can be quite robust to increases in the reverse-bias voltage. In our experiment, we abruptly increase the reverse-bias voltage from -6 V to -20 V, and observe that the structured optical spectra, the low noise RF signal, and high DC current are all maintained (Fig. \ref{Fig5}), indicating that the soliton state is preserved. Increasing the reverse-bias voltage is equivalent to reducing the effective red-detuning of the pump, which corresponds to lower soliton peak powers \cite{Herr} and causes the 3PA-induced DC current to decrease from 0.74 mA to 0.627 mA. As the reverse-bias voltage is gradually decreased from -20 V to -10 V, we observe an increase in current due to the same reasoning [Fig. \ref{Fig5}(b)]. More interestingly, we observe a hysteresis effect in the measured current, indicating that different cavity detunings are possible even for the same reverse-bias voltage. This is due to the strong dependence of FC density on dynamical intracavity power, which also makes it distinct from the thermal tuning technique. 

Another feature of electrical tuning is that the DC current is directly linked to the soliton peak power. Since the soliton peak power is determined by the pump cavity detuning and the resonator properties, we can potentially frequency stabilize the pump cavity detuning by stabilizing the DC current to a fixed value using the reverse-bias voltage. The response time of electrical tuning can be very fast due to the short FC lifetime and could be four orders of magnitude smaller than thermal tuning response time in Si$_3$N$_4$ \cite{Fainman}.

\section{Summary}
In summary, we demonstrate a modelocked silicon microresonator-based frequency comb with near-octave spanning bandwidth and low pump power requirement. We achieve a high pump-to-comb conversion efficiency where the output comb power excluding the pump mode is 40$\%$ of the total input pump power in the bus waveguide. Modelocking can be achieved without complex laser tuning techniques to overcome transient thermal instability. The electrical measurement and control of FC's in a silicon chip significantly simplifies the system design for the generation of a highly coherent and stable mid-IR frequency comb. The PIN structure enables monitoring the intracavity dynamics and tuning of the FC lifetime to produce soliton modelocking and to control and stabilize the soliton state with high bandwidth. Our approach removes the constraints of pump frequency tuning and provides a path towards a fully integrated mid-IR comb source via combining silicon microresonators with mid-IR QCL's and chip scale dual-comb spectroscopy.

\textbf{Funding.} 
Intelligence Advanced Research Projects Activity (IARPA); Defense Advanced Research Projects Agency (DARPA) (W31P4Q-15-1-0015); Air Force Office of Scientific Research (AFOSR) (FA9550-15-1-0303); National Science Foundation (NSF) (ECS-0335765, ECCS-1306035).

\textbf{Acknowledgment.} 
This work was performed in part at the Cornell Nano-Scale Facility, a member of the National Nanotechnology Infrastructure Network, which is supported by the NSF. The authors thank R.~K.~W. Lau for useful discussions.
\
%\section*{References}
% Bibliography

%Manual citation list

\end{document}